\documentclass[twoside,12pt]{conference}
\def\Chapter#1#2#3{\chapter{#1}\lhead{#3}\rhead{#2}}
\usepackage{fancyhdr}
\usepackage{epsfig}
\usepackage{graphicx}
\usepackage{multind}
\makeindex{author}
\makeindex{subject}

\def\beq{\begin{equation}}
\def\eeq#1{\label{#1}\end{equation}}
\def\eeqn{\end{equation}}

\def\beqa{\begin{eqnarray}}
\def\eeqa#1{\label{#1}\end{eqnarray}}
\def\eeqan{\end{eqnarray}}

\let\bar=\overbar

\def\O{{\cal O}}

\def\Dslash{\not{\hbox{\kern-4pt $D$}}}
\def\dslash{\not{\hbox{\kern-2pt $\del$}}}

\def\msb{{\bar{\ssstyle M \kern -1pt S}}}

\def\Dm2{\Delta m^{2}}
\def\eV2{\mbox{eV}^{2}}
\def\B8{^{8}\mbox{B}}
\def\Be7{^{7}\mbox{Be}}

\def\lapprox{\hbox{\lower .8ex\hbox{$\,\buildrel < \over\sim\,$}}}
\def\gapprox{\hbox{\lower .8ex\hbox{$\,\buildrel > \over\sim\,$}}}

\def\O16{$^{16}$O}
\def\C12{$^{12}$C}

\def\ha{\relax \ifmmode {\mbox H}\alpha\else H$\alpha$\fi}
\def\hi{\relax \ifmmode {\mbox H\,{\scshape I}}\else H\,{\scshape I}\fi}
\def\hii{\relax \ifmmode {\mbox H\,{II}}\else H\,{II}\fi}
\def\oiii{\relax \ifmmode {\mbox O\,{III}}\else O\,{III}\fi}
\def\h2{\relax \ifmmode {\mbox H\,_{2}}\else H\,$_{2}$\fi}

\def\Lsun{\relax \ifmmode {\mbox L}_\odot \else ${\mbox L}_\odot$\fi}

\begin{document}
%  FRONTMATTER:
\emptyheads   
\fancyheads
\Chapter{THE FUNDAMENTAL CYCLOTRON LINE IN 4U 1538$-$52} 
        {The CRSF in 4U 1538$-$52}{Jos\'e J. Rodes-Roca et al.}

\bigskip\bigskip

\addcontentsline{toc}{chapter}{{\it Jos\'e J. Rodes et al.}}
\label{JJRodesStart}

\begin{raggedright}  

JOS\'E J. RODES--ROCA \index{author}{Rodes-Roca, J. J.}, JOS\'E M. TORREJ\'ON \index{author}{Torrej\'on, J. M.}, J. GUILLERMO BERNAB\'EU \index{author}{Bernab\'eu, G.} \\
{\it Departament de F\'\i sica Enginyeria de Sistemes i Teoria del Senyal,
Universitat d'Alacant,
E-03080 Alacant, SPAIN}\\

\bigskip\bigskip
\end{raggedright}

{\bf Abstract:}  We present pulse phase averaged spectra of the high mass X-ray binary
pulsar \index{subject}{X-ray binaries, high mass X-ray binary} 4U 1538$-$52/QV Nor. %%@
Observations of this persistent accreting pulsar were made with the
Rossi X-ray Timing Explorer (RXTE) \index{subject}{X-ray observatory, RXTE}.

We study the variability of cyclotron resonant scattering feature (CRSF or simply
cyclotron line) in the high energy spectra of
this binary system \index{subject}{magnetic field, cyclotron %%@
resonant scattering feature}. %%@
We show that the parameters of the CRSF \index{subject}{magnetic field, CRSF}
are correlated. The first one is, as suggested
by theory, between the width and the energy of the cyclotron line. The second one is
between the relative width (defined as $\sigma_c/E_c$) and the optical depth of the
cyclotron line. %%@
We discuss these results with studies of other X-ray pulsars and their implications
on the line variability.

\bigskip

{\bf Keywords:} X-rays -- Magnetic Fields -- Accretion -- Cyclotron Lines.

\bigskip

\section{Introduction}

X-ray binary systems are a perfect astrophysical laboratory to study the behaviour
of matter and its interaction in extreme conditions of temperature, density,
gravity and its magnetic fields. These physical properties are impossible to achieve
in terrestrial laboratories. X-ray binaries are made up of a 'normal' (e. g. main
sequence) companion star and a compact object. The compact object can be either
a neutron star (NS) \index{subject}{stars, neutron star}
or a black hole \index{subject}{stars, black hole candidate}
(usually referred to as black hole candidate,
BHC). Some of the observational evidence indicate that X-ray emission is generated
by the accretion of material from the companion star onto the compact object.
In these circumstances, matter falling onto a compact star releases gravitational potential
energy, heats the matter and generates X-radiation.

In order to understand the emission properties of an accreting X-ray binary source,
we need to know:

\begin{itemize}
\item the nature of the compact object, NS or BHC; %%@
\item the strength and geometry of the magnetic field, when the compact object is
a NS;
\item the geometry of the accretion flow from the companion to the compact object;
\item the mass accretion rate;
\item the mass of the system.
\end{itemize}

Taking the mass of the companion star into account, X-ray binaries can be classified
into two main categories: High Mass X-ray Binaries \index{subject}{X-ray binaries, HMXB}
(HMXB) and Low Mass X-ray Binaries \index{subject}{X-ray binaries, LMXB} (LMXB).
These sources present a wide variety of phenomena, from quasi-periodic
oscillations (QPOs) to X-ray outbursts. However, when binary systems contain a
white dwarf (WD)\index{subject}{stars, white dwarf}
and a low mass companion star,
they are not called LMXB although X-ray emission is present. These sources are
called Cataclysmic Variables (CVs) \index{subject}{X-ray binaries,
Cataclysmic Variables} because they show very large variations in their
brightness and are also fairly faint in X-rays.

Although there are some intermediate mass X-rays binaries
\index{subject}{X-ray binaries, IMXB} (IMXB)
(e.g., XTE J1819$-$254, GRO J1655$-$40, 4U 1543$-$475),
in \cite{PRP03} they suggest that the majority of the LMXB systems may
have descended from IMXB systems.
There are some general books on
X-ray binary systems as \cite{LPH95, Longair2, FAD02, LK06} or reviews on this
topic as \cite{HMcC82, cominsky99, GS-G03} that can be useful to learn about
X-ray astronomy. In the rest of this section we only consider HMXB systems.

\begin{figure}[htb]
\begin{center}
\epsfig{file=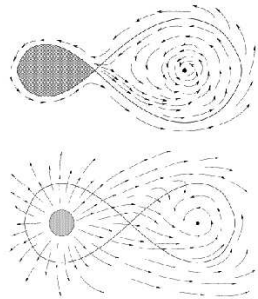,height=2in}
\caption{Mass transfer in a binary system. Top: When a massive star evolves
the outer layers begin expanding and the star
becomes a supergiant. If the star exceeds its Roche lobe, the material outside
the Roche lobe is no longer gravitationally bound to the star. This material
can now be captured by the compact object or expelled into the interstellar medium.
Bottom: When the optical companion is an O or B star, the stellar wind can be
very intense. Sometimes the compact object in a close orbit accretes material
from the dense stellar wind and causes the weak X-ray emission~\cite{shasun73}.}
\label{fig:accretion}
\end{center}
\end{figure}

In HMXB systems the companion is an O, B or Be type star with $M\geq 10 \; M_{\odot}$
and the X-ray emission is produced by capture of material from the stellar wind
by the compact object or through
Roche-lobe overflow that can also be a supplement to the mass transfer rate in HMXB
systems\cite{LPH95} (in Figure~\ref{fig:accretion} we can see an example of both
scenarios of mass transfer). Although the compact object in HMXBs may be a black hole,
as in the case of Cyg X-1 (the first X-ray source discovered in the constellation
Cygnus), they usually contain a NS \index{subject}{compact object, neutron star}.

Most HMXBs fall in one of the two main subgroups: those in which the primary has
evolved away from the main sequence and becomes a supergiant (SG/X-ray binary)
\index{subject}{X-ray binaries, SG/X-ray binary} and those in which the primary
has not reached the supergiant phase and is characterized by Balmer
and HeI emission lines
(Be/X-ray binary) \index{subject}{X-ray binaries, Be/X-ray binary}.

In general, SG/X-ray binary systems \index{subject}{X-ray binaries, SG/X-ray binary}
are sources of persistent X-ray emission \index{subject}{persistent X-ray emission}.
At some stage of the binary evolution, the supergiant star ejects much of its
mass in the form of a stellar wind; the NS \index{subject}{star, neutron star}
can pick up some of these particles gravitationally and becomes a weak X-ray source.
But, in the course of star evolution, another possibility may be an increase in
radius of the companion star becoming a supergiant; then the outer layers of its
envelope escape from the Roche lobe of the supergiant star through the internal
Lagrangian point and becomes a bright X-ray source.

\begin{figure}[htb]
\begin{center}
\epsfig{file=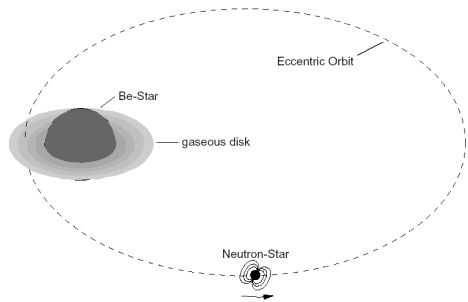,height=2in}
\caption{The NS revolves around the Be star in an eccentric orbit.
When the NS comes inside the circumstellar disk of the Be star,
accretion takes place and causes an X-ray outburst. However, while the NS is
far away from the Be star, it is in quiescence and there is no X-ray emission.}
\label{fig:bexrb}
\end{center}
\end{figure}

The most numerous class of HMXBs are Be/X-ray binaries, called also hard X-ray
transients (see the catalog of \cite{Liu2000}).
The companion star is an Oe or Be star, which are O or B stars with
bright optical emission lines originated by circumstellar disc. It is believed that
this envelope around the Be star is caused by its fast rotation. The orbit of
the NS around the Be star is usually eccentric, so when is far away from the companion
star cannot accrete material from this envelope. When the NS approaches
periastron, it will be able to accrete material and the observer sees an
X-ray outburst. Figure \ref{fig:bexrb} shows a collapsed object in an eccentric
orbit around a Be star (for a review of Be/X-ray binaries, \cite{ignacio98, janusz02}).

However,
a new kind of HMXBs was discovered by INTErnational Gamma Ray Astrophysical Laboratory
(INTEGRAL) \index{subject}{X-ray observatory, INTEGRAL}, called Supergiant Fast X-ray
Transients \index{subject}{X-ray binaries, SFXT/X-ray binary} \cite{ignacio06},
which are characterized by the occurrence of very fast X-ray outbursts.
In \cite{ignacio06} they have shown that at least a significant fraction of
them are associated with supergiant stars.

Accreting X-ray pulsars \index{subject}{X-ray binaries, X-ray pulsars} were discovered
by Giacconi et al. in 1971~\cite{giacconi71} when they %%@
discovered the existence of periodic pulsations in the X-ray emission from
Centaurus X-3. X-ray pulsars are rotating, highly magnetized NS %%@
accreting material from a companion binary star \cite{WSH83, nagase89, bildsten97,
staubert03}.
Most known accreting X-ray pulsars belong to the HMXB class, such as 4U 1538$-$52/QV Nor. %%@
For accreting binary pulsars we measure their magnetic fields through the
presence of cyclotron resonance scattering features (CRSFs) \index{subject}{magnetic
fields, cyclotron resonance scattering features} in their X-ray spectra because
the cyclotron energy ($E_{cyc}$) and the magnetic field strength (B) are related to each
other as $E_{cyc} = 11.6 \; B(10^{12} \; G) \; (1+z)^{-1}$ keV, where z is the
gravitational redshift.

\section{Observational data}

RXTE \index{subject}{X-ray observatory, RXTE} observed 4U 1538$-$52 between 1996
November 24 and 1997 December 13.
To obtain the exact orbital phase we used the best fit
orbital ephemeris from \cite{makishimaEPH}. The exact time and the orbital
phase are listed in \cite{rodesPhD}.

In our analysis we used data
from both RXTE pointing instruments, the Proportional Counter Array (PCA)
and the High Energy X-ray Timing Experiment (HEXTE). To extract the
spectra, we used the standard RXTE \index{subject}{X-ray observatory, RXTE}
analysis software FTOOLS.

The PCA consists of five co-aligned Xenon proportional counter units with
a total effective area of $\sim$6000 cm$^2$ and a nominal energy range
from 2 keV to over 60 keV (\cite{jahoda}). However, due
to response problems above $\sim$20 keV and the Xenon-K edge around
30 keV, we restricted the use of the PCA to the energy range from 3 keV
to 20 keV (see also \cite{ingo}).

The HEXTE consists of two clusters of four NaI(Tl)/CsI(Na) Phoswich
scintillation detectors with a total net detector area of 1600 cm$^2$.
These detectors are sensitive from 15 keV to 250 keV
(\cite{hexte}). However, response matrix, instrument background and source
count rate, limit the energy range from 17 to 100 keV. Background
subtraction in HEXTE is done by source-background swapping of the two
clusters every 32 s throughout the observation. In order to improve the
statistical significance of the data, we added the data of both HEXTE
clusters and created an appropriate response matrix by using a 1:0.75
weighting to account for the loss of a detector in the second cluster.
We also binned several channels together of the HEXTE data at higher
energies and chose the binning as a compromise between increased
statistical significance while retaining a reasonable energy resolution.

\section{X-ray spectral analysis}

The X-ray spectrum in accreting X-ray pulsars has been investigated during
the last two decades. Nevertheless, there still exists no convincing
theoretical model for the continuum of this kind of sources
(\cite{harding}, and references therein). Therefore, we have to use
empirical models of the continuum in the fitting process. In the RXTE
\index{subject}{X-ray observatory, RXTE}
energy band these models take the general form of
a power law times an exponential above a characteristic
cutoff energy.

\subsection{Continuum model}

We achieved a good description of the continuum X-ray spectra of
this source using the standard pulsar continuum shape and customize %%@
models from {\sc xspec} \cite{rodesPhD}. All of them are modified by
a photoelectric absorption at low energies, a fluorescence iron emission
line at 6.4 keV and the fundamental CRSF at 20 keV discovered by Ginga
\index{subject}{X-ray observatory, Ginga}. The description of the continuum
by physical models gave parameters that were not acceptable.
We also modeled the observational data with several other
standard pulsar continuum but they did not describe the spectra properly
in the 7$-$16 keV energy band \cite{RTBII}.
Therefore, in this paper we described the continuum produced
in the accretion column of the NS by the Negative Positive power laws
EXponential ({\sc npex}) component modified by previous features. It was
introduced by \cite{mihara} and given by the formula: 
\begin{equation}
	NPEX(E) = A \cdot \left( E^{-\Gamma_1} + B \cdot E^{+\Gamma_2} \right)
      \cdot e^{-E/E_{fold}} \,, 
\end{equation}
where $\Gamma_1$ and $\Gamma_2$ are positive and $E_{fold}$ is the folding
energy of the high energy exponential cutoff. We used a Gaussian emission line
at $\sim$6.4 keV due to the fluorescence iron line and a multiplicative factor
to fit the cyclotron line of the form:  
\begin{equation}
  cyclabs(E) = \exp \left( \frac{-\tau_c \cdot \left[ \frac{\sigma_c \cdot E}
         {E_c} \right]^2}{(E-E_c)^2+\sigma_c^2} \right) \,,
\end{equation}
where $\tau_c$, $\sigma_c$ and $E_c$ are the optical depth, the width and the
energy of the CRSF, respectively. We found that {\em cyclabs} model provides
reasonable fits to the data. The observed X-ray spectrum of 4U 1538$-$52 is
modified by photoelectric absorption due to the stellar wind of QV Nor.

\subsection{CRSF variability \index{subject}{CRSF, variability}}
\label{CRSFparameters}

The main aim of this paper is to study the variation of the parameters of the
cyclotron absorption line \index{subject}{cyclotron line} and their relations.
The CRSF centered at $\sim$20 keV varies by $\sim$15\% through the pulse
\cite{clark90}. In our pulse phase averaged spectra the
parameter identified with the magnetic field, $E_c$, does not depend
significantly on the orbital phase as is expected because it
originates in the polar caps of the NS or the accretion column
where the X-rays and the pulse forms \cite{rodesPhD}. In fact, the energy of the CRSF
only varies by $\sim$4\% (in some cases the same that the uncertainties
of the parameter at 90\% confidence level). Therefore, the stellar wind
does not modify the accretion onto the NS during an orbital period
significantly. However, we found that the relative width of the CRSF and its
optical depth \index{subject}{CRSF, optical depth} are
correlated, as well as the CRSF energy and its width \index{subject}{CRSF, width}.

Although pulse phase resolved spectroscopy \index{subject}{pulse phase spectroscopy} 
allows us to study the variation of the
pulsar emission over the X-ray pulse, it seems that the correlations between
the parameters of the CRSF are indeed real in pulse phase averaged spectroscopy.
Using a consistent set of models \cite{coburn2} Coburn et al. parameterized
a large sample of X-ray pulsars observed with RXTE which exhibit
cyclotron lines. They found a new correlation between the relative width of
the CRSF, $\sigma_c/E_c$, and its optical depth, $\tau_c$. Although they used
pulse phase averaged spectra, another study of GX 301$-$2
\cite{ingophd} with pulse phase resolved spectra confirmed this correlation.
They also reported another correlation between the cyclotron line width
 \index{subject}{CRSF, width} 
and the energy of the cyclotron line, both in pulse phase average and resolved
spectra. Therefore we have plotted our fitted results for the cyclotron
line parameters to check these correlations.

\begin{figure}[htb]
\begin{center}
\epsfig{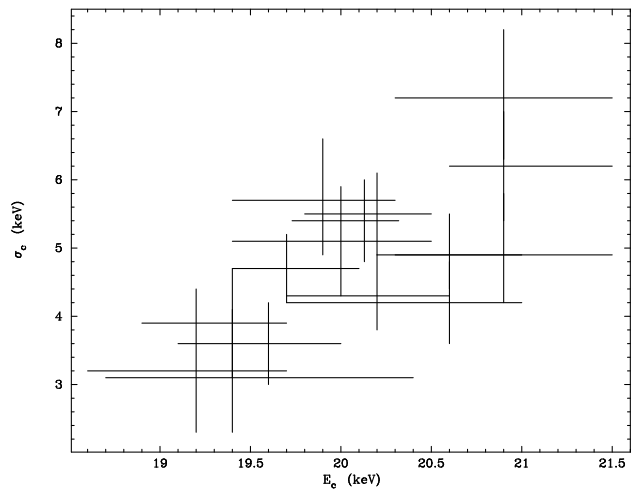}
\caption{CRSF width $\sigma_c$ versus CRSF energy $E_c$ in 4U 1538$-$52.
      Note the moderate correlation between them. The bars
      indicate the uncertainties at 90\% confidence level.}
\label{correlac1}
\end{center}
\end{figure}

In Figure~\ref{correlac1} we show the first correlation between the
CRSF parameters in the $\sigma_c - E_c$ plot. For a self-emitting atmosphere,
the cyclotron line width and his energy is given by \cite{meszaros1}:
   \begin{equation}\label{EcWc}
      \sigma_c \approx E_c \,  \left( 8 \, \ln 2 \, \frac{k \, T_e}{m_e \, c^2}
      \right)^{\frac{1}{2}}
      \, \left| \cos \theta \right| \,,
   \end{equation}
where $k \; T_e$ is the temperature of the electrons along the magnetic field
lines, $E_c$ the cyclotron line energy, and $\theta$ the viewing angle with
respect to the magnetic field. Therefore a linear correlation is only possible
if $\cos \theta$ does not change significantly. Our results imply an angle
close to zero when $k \; T_e$ is 5.2 keV and can take a value of 18$^o$
if the energy of electrons is 5.8 keV.

\begin{figure}[htb]
\begin{center}
\epsfig{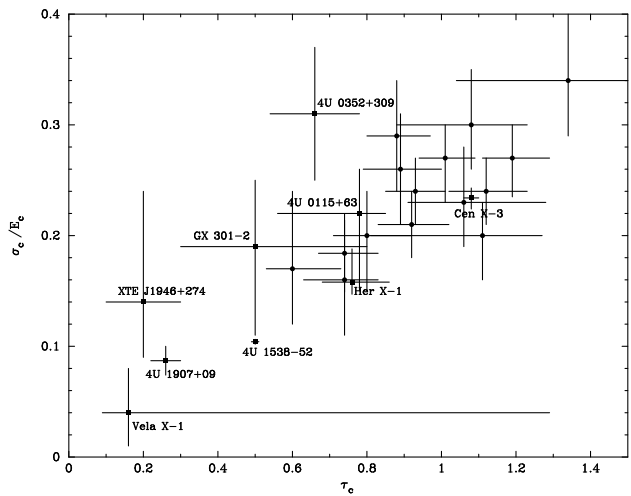}
\caption{Fractional CRSF width $\sigma_c/E_c$ versus the optical
depth of the CRSF for several accreting pulsars from RXTE. Filled squares:
values derived by \cite{coburn2} from phase averaged
spectra. Filled circles: values derived from phase averaged spectra for
4U 1538$-$52 (this work, the bars
indicate the uncertainties at 90\% confidence level). All quantities
refer to the pulse phase averaged values. Note the moderate correlation between them.}
\label{correlac2}
\end{center}
\end{figure}

The parameter $E_{fold}$ in the {\sc npex} model is the typical temperature of the
X-ray emitting plasma in keV. Basic Comptonization theory suggests that
$k \; T_e$ can be estimated from the folding energy of the pulsar
continuum. Furthermore, if we assume that the seed photons for the
Compton scattering in the accretion column are created throughout the
volume of the accretion column, then detailed Monte Carlo simulations
show that the optical depth of the CRSF \index{subject}{CRSF}
 is expected to be largest when the line
of sight is almost perpendicular to the direction of the magnetic field
\cite{isenberg}. If the temperature in the accretion
column is constant, these models predict an anti correlation between
the optical depth and the relative width of the CRSF. As we show in Figure~\ref{correlac2},
our phase averaged spectra results for 4U 1538$-$52 indicate a moderate
correlation opposite to the models, and the conclusion is that the
temperature in the accretion column is not constant. This correlation
indicates that as CRSF increase in optical depth \index{subject}{CRSF, optical depth},
the CRSF relative width increases
as well.

In the above discussion, it has been assumed that the X-ray emission is
produced from one homogeneous emission region. However, this source
has contributions from both magnetic polar caps, which could influence
the observed correlation. In fact, in 4U 1538$-$52 the two polar caps are
observed and are unequal and non antipodal \cite{bulik}.
Therefore we can expect the
parameters of the X-ray continuum emitted by each pole to be different,
which can be reflected by changes in the observed continuum parameters.
The variation in the folding energy of this system could explain this
correlation caused by a mixture flux from the two polar caps. However,
this reason cannot be the only one because, for example, in GX 301$-$2
the folding energy does not change significantly \cite{ingophd}.

\begin{figure}[htb]
\begin{center}
\epsfig{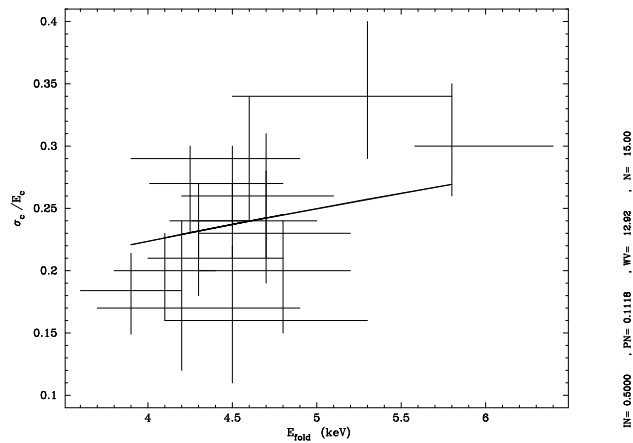}
\caption{Fractional CRSF width $\sigma_c/E_c$ versus the folding energy. All quantities
refer to the pulse phase averaged values. The bars
indicate the uncertainties at 90\% confidence level.}
\label{correlac3}
\end{center}
\end{figure}

Assuming that equation~\ref{EcWc} is correct and $\cos \theta$ does not
change appreciably, we expect a relationship between the folding energy
and the relative width of the CRSF. As we can see in Figure~\ref{correlac3},
we found that the relationship was consistent with a power law
$\sigma_c/E_c \propto E_{fold}^{0.5}$, indicating little variation of
the angle $\theta$ and according to equation~\ref{EcWc}.
However, the statistic of the data is such that,
from Figure~\ref{correlac3}, we cannot
distinguish between the last relationship and a linear correlation
$\sigma_c/E_c \propto E_{fold}$, at a statistically significant level.

\section{Summary and conclusions}

We have studied the variability of the parameters of the fundamental cyclotron absorption line in the
HMXB 4U 1538$-$52. In order to explain this variability, we have also studied the relationship among
the continuum ($E_{fold}$) and cyclotron line parameters. %%@

As shown in Figure~\ref{correlac2}, our pulse phase averaged results for 4U 1538$-$52
are in agreement with the correlation found by \cite{coburn2}. Also the
pulse phase resolved results for GX 301$-$2 \cite{ingophd} noticed the
same correlation, so it suggests that it is not due to effects of averaging.
In terms of the relativistic cross sections, this result is in the opposite
sense.

Furthermore, the fundamental CRSF of this source has a width nearly proportional
to its energy. If we consider only a thermal broadening, then the variation
of the viewing angle is not significant.

Our third result is a relationship between the folding energy $E_{fold} = k \; T_e$ and the
relative width of the cyclotron line. Figure~\ref{correlac3} shows this correlation
and it implies viewing angle close to zero which is consistent with the previous
correlation.

Our results for the spectral analysis of the RXTE data of 4U 1538-52 can be
summarized as follows:

\begin{itemize}

\item The absorbed {\sc npex} continuum model provides a reasonable description
of all the spectra used in this research. It approximates a photon number spectrum for
an unsaturated thermal Comptonization in a plasma of temperature $T_e$.

\item The correlation between the width of the fundamental CRSF and its
energy suggests little variations of the angle between the line of sight
and the magnetic field in the accretion column.

\item The relationship between the folding energy and the relative width
of the fundamental CRSF implies that $\cos \theta$ is close to 1. As an
example, $\cos 26^o \sim 0.9$, so a little variation
of the viewing angle and its relationship is consistent.

\item The correlation between the relative width of the cyclotron line
and its optical depth implies changes in the angle between the line of sight
and the magnetic field at the NS poles, but in the opposite
sense to the numerical simulations for the optical depth of the line.

\end{itemize}

\bigskip
{\it Acknowledgements:} This research is partially supported by the Spanish
\emph{INTEGRAL: Observaciones multifrecuencia de sistemas binarios
      de rayos X} project number ESP2001-4541-PE and \emph{International
      Gamma Ray Astrophysics Lab. Operaciones. C3} project number
      ESP2002-04124-C03-03. This research has made use of data obtained
      through the High Energy Astrophysics Science Archive Research
      Center Online Service, provided by the NASA/Goddard Space Flight
      Center.

%\end{document}

%%
%%
%%
%%%%%%%%%%%%%%%   Author and Subject Index
%\printindex{author}{Author Index}
%\blankpage
%\printindex{subject}{Subject Index}
%\blankpage
\end{document}